\documentclass{bmcart}

\RequirePackage{natbib}
\RequirePackage{hyperref}
\usepackage[utf8]{inputenc} 
\usepackage{cleveref}
\usepackage{graphicx}
\usepackage{caption}
\usepackage{subcaption}
\usepackage{makecell}

\startlocaldefs
\endlocaldefs

\begin{document}

\begin{frontmatter}

\begin{fmbox}
\dochead{Research}


\title{Adaptation of Student Behavioural Routines during COVID-19: A Multimodal Approach}


\author[
   addressref={aff1, aff2},                   
   corref={aff1},                       
   email={ngirardini@fbk.eu, nicolo.girardini@unitn.it}   
]{\inits{NAG}\fnm{Nicolò Alessandro} \snm{Girardini}}
\author[
   addressref={aff1},
   email={centellegher@fbk.eu}
]{\inits{SC}\fnm{Simone} \snm{Centellegher}}
\author[
   addressref={aff2},
   email={andrea.passerini@unitn.it}
]{\inits{AP}\fnm{Andrea} \snm{Passerini}}
\author[
   addressref={aff3},
   email={ivano.bison@unitn.it}
]{\inits{IB}\fnm{Ivano} \snm{Bison}}
\author[
   addressref={aff2},
   email={fausto.giunchiglia@unitn.it}
]{\inits{FG}\fnm{Fausto} \snm{Giunchiglia}}
\author[
   addressref={aff1},
   email={lepri@fbk.eu}
]{\inits{BL}\fnm{Bruno} \snm{Lepri}}


\address[id=aff1]{
  \orgname{Fondazione Bruno Kessler (FBK)}, 
  \street{Via Sommarive 18},                     %
  \postcode{38123}                                
  \city{Trento},                              
  \cny{Italy}                                    
}
\address[id=aff2]{%
  \orgname{University of Trento, Department of Information Engineering and Computer Science},
  \street{Via Sommarive 9},
  \postcode{38123}
  \city{Trento},
  \cny{Italy}
}
\address[id=aff3]{%
  \orgname{University of Trento, Department of Sociology and Social Research)},
  \street{via Verdi 26},
  \postcode{38122}
  \city{Trento},
  \cny{Italy}
}


\begin{artnotes}
\end{artnotes}

\end{fmbox}


\begin{abstractbox}

\begin{abstract}
One population group that had to significantly adapt and change their behaviour during the COVID-19 pandemic is students. While previous studies have extensively investigated the impact of the pandemic on their psychological well-being and academic performance, limited attention has been given to their activity routines. In this work, we analyze students' behavioural changes by examining qualitative and quantitative differences in their daily routines between two distinct periods (2018 and 2020). Using an Experience Sampling Method (ESM) that captures multimodal self-reported data on students' \textit{activity}, \textit{locations} and \textit{sociality}, we apply Non-Negative Matrix Factorization (NMF) to extract meaningful behavioural components, and quantified the variations in behaviour between students in 2018 and 2020. Surprisingly, despite the presence of COVID-19 restrictions, we find minimal changes in the activities performed by students, and the diversity of activities also remains largely unaffected. Leveraging the richness of the data at our disposal, we discovered that activities adaptation to the pandemic primarily occurred in the \textit{location} and \textit{sociality} dimensions.

\end{abstract}




\end{abstractbox}
%

\end{frontmatter}



\section{Introduction}
\label{sec:intro}

The COVID-19 pandemic has had a wide-ranging impact on various components of our daily lives. The implementation of government restrictions such as travel bans, business and school closures, stay-at-home orders and physical distancing mandates aimed at preventing the spread of the virus \cite{perra2021non, brauner2021inferring, covid_spread}. These measures resulted in substantial changes in how we live, work, and socialize with long lasting impacts, among others, on human mobility and encounters \cite{covid_spread, living_pandemic, urban_behav}, employment \cite{adams2020inequality, baek2021unemployment}, education \cite{van2020covid, learning_review, yang2021college}, mental health and well-being \cite{wellbeing, fatigue, brodeur2021covid}.

Traditionally, social scientists have relied on qualitative techniques such as direct observation and fieldwork to study various aspects of our daily lives and gain insights into these social phenomena \cite{goffman1959, brown2014}. However, these methods have limitations as they can be time-consuming, resource-intensive, and challenging to replicate, making them less suitable for large-scale studies.

In the last decades, the emergence of the digital sensing revolution has provided researchers with vast amounts of quantitative data generated by, among others, social media platforms, credit cards, and mobile phones. Digital data sources collected with these technologies offer unprecedented opportunities to gain insights into individual and group behaviours with a breadth and depth that was previously inconceivable \cite{lazer2009computational, lazer2020}. This wealth of digital data has become particularly valuable during the COVID-19 pandemic, enabling researchers to uncover patterns and trends in human behaviour and to understand the changes that have occurred in our daily lives and society as a whole \cite{hou2021intracounty, bonaccorsi2021socioeconomic, persson2021monitoring}.

While these data sources allow us to track the behaviour of thousands or even millions of individuals, it's important to note that the temporal resolution of the collected data is not continuous but rather tied to specific events (e.g., phone calls or traffic data for mobile phone data, app usage for GPS location data, credit card usage for spending data, etc.). Furthermore, these data sources often represent only one aspect of human behaviour at a time such as communication, spending, and mobility.

To overcome these limitations researchers have implemented ``living labs" studies, which collect multiple sources of data \cite{reality_mining, stopczynski2014measuring, centellegher2016mobile, wenet, madan2011sensing, students_life, aharony2011social}. In these studies, participants were equipped with mobile phones and/or sensing apps installed on their devices to collect data from diverse sources and sensors, including phone and SMS logs, Bluetooth proximity interactions, GPS location data, accelerometer data, etc. Additionally, surveys and experience sampling methodologies were employed to gather information on individual characteristics (e.g., attitudes, personality traits, etc.), on states (e.g., mood, stress levels, etc.) as well as situational and activity data. By integrating data from multiple sources, living labs deployments offer the advantage of capturing a richer representation of the individuals under study.

In our work, we leverage an Experience Sampling Method (ESM) \cite{csikszentmihalyi2014experience} approach on smartphones to capture self-reported information on the daily activities of university students. ESM is a methodology aiming at collecting information on the behaviours and emotions of study participants throughout their daily activities and routines \cite{raento2009smartphones, pejovic2016, van2017experience}. As in traditional time diary studies, ESM collects data by means of study participants’ self-reports. However, participants, unlike in diary studies, are proactively triggered at various moments during the day. In our work, we ask the study participants to report information on their \textit{activities} with a prompt every 30 minutes, and we collect not only the activity of the students but also their \textit{location} and \textit{sociality} (i.e., with whom they were). Hence, this approach produces a unique understanding of students' behaviour from the point of view of these three different behavioural dimensions. Moreover, the ESM-based data collection happened in two periods, in 2018 and 2020, and we leverage these data to analyse how students' routines changed across all these three behavioural dimensions during the COVID-19 pandemic.  

In particular, we use \emph{Non-Negative Matrix Factorization} (NMF) \cite{sra2005generalized, nmf} to extract meaningful behavioural components (i.e., routines) from self-reported data. By analyzing the extracted components and their associated weights, we are able to quantify the differences in behaviour between students in 2018 and students in 2020. Our findings indicate that despite the impact of the pandemic, the \textit{activity} routines of students in 2018 and 2020 remained largely similar with minimal differences observed. However, with the multimodal information collected through the time diaries, we discovered that the behavioural dimensions of \textit{location} and \textit{sociality} exhibited stronger signals of adaptation. To delve deeper into this shift of behaviour, we first examine the relationship between \textit{activities} and \textit{locations}, then the relationship between \textit{activities} and \textit{sociality}. This approach allows us to identify the activities that were associated with diverse \textit{locations} and \textit{sociality} routines before and during the pandemic. Notably, activities such as studying and attending lessons, which had a social component before the pandemic, transitioned to solitary activities, done in private homes, even if students could go to the university. We find that most of the extracted routines in 2020 follow this pattern, and take place at home and alone, while in 2018 students visited more diverse locations and spent more time with friends. For example, social life is more present in 2018, with students spending time with friends, while in 2020 this activity is done with roommates, thus people residing in the same household.

\section{Related Work}
\label{sec:rel_work}

\subsection{Behaviours and Routines Modelling}
\label{subsec:behave_route_mod}

In recent years, the field of modelling human behaviour and routines has witnessed significant growth, thanks to advancements in mobile sensing \cite{lane2010survey} and experience sampling approaches \cite{raento2009smartphones}.

Various techniques have been employed to understand individual behaviour. Principal Component Analysis (PCA) was used to extract students' routines (\textit{eigenbehaviours}) \cite{eigen}, from the Reality Mining dataset \cite{reality_mining}. Techniques typically used for Natural Language Processing tasks \cite{yao2009efficient} have been used to extract regular behavioural patterns from sequential data, such as Latent Dirichlet Allocation \cite{blei2003latent}, which was also used to group individual activities into routines \cite{tm_discovery, soc_geo_lda}. Other sequence analysis techniques have been used as well \cite{task_footprint, everyday_patterns}, from modelling purchasing behaviour \cite{purch_cred_card} to the analysis of individual mobility \cite{living_pandemic}.

Other methods include more complex tools, such as Hidden Markov Models \cite{hmm}, Conditional Random Fields \cite{crf} and Bayesian Networks \cite{dynamic_bn}. These approaches are commonly used to build predictive and generative models, and can be used to detect the activities from sensor data and images \cite{crf, dynamic_bn} or to model cyclic behaviour and its anomalies \cite{hmm}. 
While these methods have their advantages, in our study, we aim to identify easily interpretable and human-readable patterns to compare and measure behavioural differences.

In recent years, \textit{Non-Negative Matrix Factorization} (NMF) \cite{sra2005generalized} has gained popularity, thanks to its interpretability. NMF has proven to be effective in identifying repeating patterns in the daily behaviours of people, both at an individual and a collective level. NMF has been successfully applied in various domains, such as in analyzing social interactions in networks \cite{gauvin2014detecting}, understanding of chronotypes \cite{nmf_rythms}, and identifying routines in urban environments \cite{purch_cred_card,nmf_latent}. Inspired by these studies, we use NMF in our methodology, as it allows us to associate the resulting patterns with realistic individual behaviour, and the weights associated with these patterns provide a quantitative measure of behavioural differences. Additionally, following \cite{behav_context}, we include in our methodology, the use of multiple modalities to analyse subjects' behaviour, which includes (i) activity, (ii) location, and (iii) sociality dimensions.

\subsection{Behavioural Change and the Pandemic}
\label{subsec:behave_change_covid}

There has been an upsurge of scientific research in response to the COVID-19 crisis aimed at understanding and mitigating the spread of the virus \cite{covid_spread, covid_modelling}. The various consequences and effects of COVID-19 restrictions on the population, spanning economic, social, and psychological domains have been tackled in diverse disciplines.

From an economic perspective, studies have demonstrated that the effects of restrictions differ across individuals with varying income levels, often increasing segregation. For instance, in Italy, the mobility reduction was higher in high fiscal capacity municipalities, which also display high income inequality \cite{economic}. Similar findings have emerged in other countries such as France and the United Kingdom, where the mobility network became more fragmented on a national scale but maintained more connections on a small scale \cite{mobility_fui}. In the United States, researchers have observed a decrease in encounters between citizens of different income levels in urban areas, hindering economic recovery and city growth \cite{urban_behav}, despite mobility returning to pre-pandemic levels.

Another area of interest has been the well-being of the population, with a particular focus on students, enhancing research that was present even before the pandemic \cite{depression}. The impact of the pandemic on students' educational paths has been substantial \cite{learning_review}, as shown in studies on academic performance and mental well-being \cite{covid_performance}. Research has indicated a significant increase in stress and anxiety levels among younger students during lockdowns \cite{students_activity}, particularly college and university students \cite{students_stress, yang2021college, students_mental}.

Previous studies on students primarily rely on surveys and self-reported data, often neglecting the impact on behavioural routines. Furthermore, only a few studies consider multiple dimensions of behaviour and well-being with high-resolution data. For instance, the study conducted in \cite{students_life} combines surveys with mobile sensing data to explore how students' perspectives and concerns regarding the pandemic lead to diverse responses to restrictions and shifts in behaviour.

In contrast, our work demonstrates the utility of Experience Sampling Methods on smartphones, which provide a wealth of information on activities, locations, and social interactions in a real-world setting. These methodologies offer a level of granularity, particularly in activity definition, that cannot be achieved through passive sensor data collected by mobile phones. Moreover, we highlight how these methodologies can be harnessed to understand multiple dimensions of students' behaviour, both independently and in conjunction with one another.

\section{Data}
\label{sec:data}

The data utilized in this work are derived from two separate datasets collected during two distinct living lab studies: the \textit{SmartUniTn} study \cite{smartunitn, giunchiglia2017mobile, giunchiglia2018mobile} conducted in 2018, and the subsequent \textit{WeNet} study \cite{wenet, assi2023complex, generalization} conducted in 2020.

The \textit{SmartUniTn} study gathered both sensor data and questionnaire responses from students at the University of Trento, employing the \textit{iLog} \cite{ilog} smartphone application. Over a period of two weeks, with a frequency of 30 minutes, participants were requested to complete a \textit{Time Diary} via the Experience Sampling Method (ESM) application. Each entry in the diary required students to report their \textit{activity} (e.g., studying, eating, etc.), \textit{location} (e.g., home, university, bar, etc.), and \textit{sociality}, indicating whether they were alone or in the company of others (e.g., alone, friends, relatives, etc.). The response options were predefined to ensure uniformity and minimize user biases in activity descriptions (refer to \cite{smartunitn} for the available sets of responses). This comprehensive dataset, which was collected from May 10th to May 23rd, 2018, provides a multimodal representation of student behaviour in a period when students attended classes and prepared for the summer exam session.

The subsequent \textit{WeNet} study \cite{wenet} expanded upon the \textit{SmartUniTn} study by extending the data collection to include seven universities worldwide, including the University of Trento. To ensure consistency with the previous study, we only focused on the data from the latter university. The data collection process closely followed that of the \textit{SmartUniTn} study, with refinements made to the sets of activities and locations (for detailed information, see \cite{wenet}). Notably, the data collection occurred during the COVID-19 pandemic, and as such, specific government restrictions were in place at the time. The Time Diaries were collected between November 14th and November 30th, 2020, a period during which the government of the province of Trento implemented several restrictions \cite{covid_tn}. Certain activities, such as gyms, were closed, while others, like shops, had limited access. Distancing measures were enforced in various contexts, including bars and restaurants, where in addition to physical distancing, time restrictions were imposed on the duration people could spend at these locations. Moreover, while schools were generally closed, first-year students participating in our study could attend lessons in-person.

\section{Methods}
\label{sec:methods}

\subsection{Data Processing}
\label{subsec:data_process}

Since the data collection of the two studies was performed in different iterations, the student samples are different. However, the study settings remained consistent, and the population under examination was similar as both studies focused on students. To facilitate a more robust comparison between the groups of students in 2018 and 2020, we performed a mapping between the categories in the \textit{Time Diaries} to ensure the comparability of activities, locations, and sociality categories. Given that the data collected in 2020 were more refined, the mapping primarily involved the alignment of certain activities and locations, where multiple entries from 2020 were mapped into a single entry from 2018. For instance, activities like \textit{Free Time Study}, \textit{Arts}, \textit{Hobbies}, and \textit{Games} in 2020 were all mapped to \textit{Hobbies} as in 2018. A comprehensive list of activity and location mappings can be found in \Cref{app:mapping}.

Subsequently, we organized the data to follow circadian cycles starting at 5 AM, ensuring that the daily Time Diaries of each student were appropriately aligned. We excluded days and students with insufficient reported data. Additionally, as our focus was on typical behaviour during a regular student week, we excluded weekends from our analysis.

Given that Time Diaries provide insights into three different facets of behaviour (\textit{activity}, \textit{location} and \textit{sociality}), we created three distinct matrices, each corresponding to the type of information collected. In these matrices, the rows represent students, while the columns represent their behaviour throughout the day. To achieve this, we allocated a number of columns (or features) equal to $c = |A|*48$, where $A$ represents the set of possible choices for each data type (e.g., the possible activities for the activity matrix), and \textit{48} represents the number of time intervals within a day (given that the data was collected every 30 minutes). Consequently, the matrices have dimensions of $N \times c$, with $N$ denoting the \textit{247} students from both datasets. Each matrix contains the normalized counts of observed behaviour (\textit{activity}, \textit{location} and \textit{socialty}) within specific time slots for each student.

\subsection{Non-Negative Matrix Factorization}
\label{subsec:nmf}

To extract meaningful daily routines from our data, we applied \textit{Non-Negative Matrix Factorization (NMF)} \cite{sra2005generalized, nmf}, a matrix decomposition technique. The algorithm\footnote{The algorithm was implemented using the \textit{scikit-learn} Python package}, identifies recurring patterns within the input data and approximates the original data by representing it as a linear combination of these patterns. NMF operates on non-negative matrices and decomposes them into two non-negative matrices, making the results easier to interpret, as applications in image analysis \cite{nmf_example1} and text mining \cite{pauca2004text} show.

In more detail, NMF decomposes each of our behavioural matrices, denoted as $X$, into an approximation $X \approx WK$. Here, $K$ represents the $r \times c$ matrix of extracted components, where $c$ denotes the number of features defining the behaviour (as defined in \Cref{subsec:data_process}), and $r$ represents the number of extracted components or \textit{rank}. Consequently, the matrix $K$ contains the weights of each feature for each component, while the matrix $W$ has dimensions $N \times r$ and contains the weights of the components for each student. The algorithm leverages its inherent clustering property to identify common observations within $X$, storing the discovered repeating patterns in $K$. For example, in the $N \times c$ \textit{location} matrix, the matrix of extracted components $K$ captures common textit{location} routines among students, while the matrix $W$ contains the weights of the extracted components for each student. The original behaviour of a student $i$ can thus be reconstructed as: $x_{i} = w_{i1} k_{1} + w_{i2} k_{2} + \ldots + w_{ir} k_{r}$.

The approximation $X \approx WK$ is achieved by minimizing the reconstruction error computed using the \textit{Frobenius distance}:
\begin{equation}
    E = || X - WK ||_{F} = \sqrt{\sum_{i,j} \left( x_{i,j} - \sum_{r} w_{i,r} k_{r,j}\right)^2}
\end{equation}

The NMF algorithm takes the rank $r$ as a parameter, indicating the number of components to extract. To determine the optimal number of components, we used the \textit{cophenetic correlation coefficient} \cite{cophenetic}, a common practice in the literature \cite{nmf_rythms,urban_behav}. The coefficient measures the preservation of the distance between different observations during the NMF transformation, indicating the reliability of the components. Typically, the rank is chosen where the cophenetic coefficient is maximized and begins to decline. In our analysis, we selected the number of components by considering the cophenetic coefficient, and when the choice was not evident, we prioritized the rank $r$ that yielded the most interpretable components.

\section{Results}
\label{sec:res}

\subsection{Investigating Students' Behaviour in 2018 and 2020}
\label{subsec:stud_behav}

We started our analysis by examining the behavioural patterns present in both samples of students in 2018 and 2020. We applied NMF, as described in \Cref{sec:methods}, on the matrices with students from both the 2018 and 2020 data collection campaigns. Including both samples allowed us to analyze the behaviour of the two groups of students and to extract the most dominant students' routines in both 2018 and 2020. By using the three separate matrices that represented \textit{activity}, \textit{location}, and \textit{sociality} behaviours, we identified 6 \textit{activity} routines, 4 \textit{location} routines, and 7 \textit{sociality} routines.

\begin{figure}[ht!]
    \includegraphics[width=.98\textwidth]{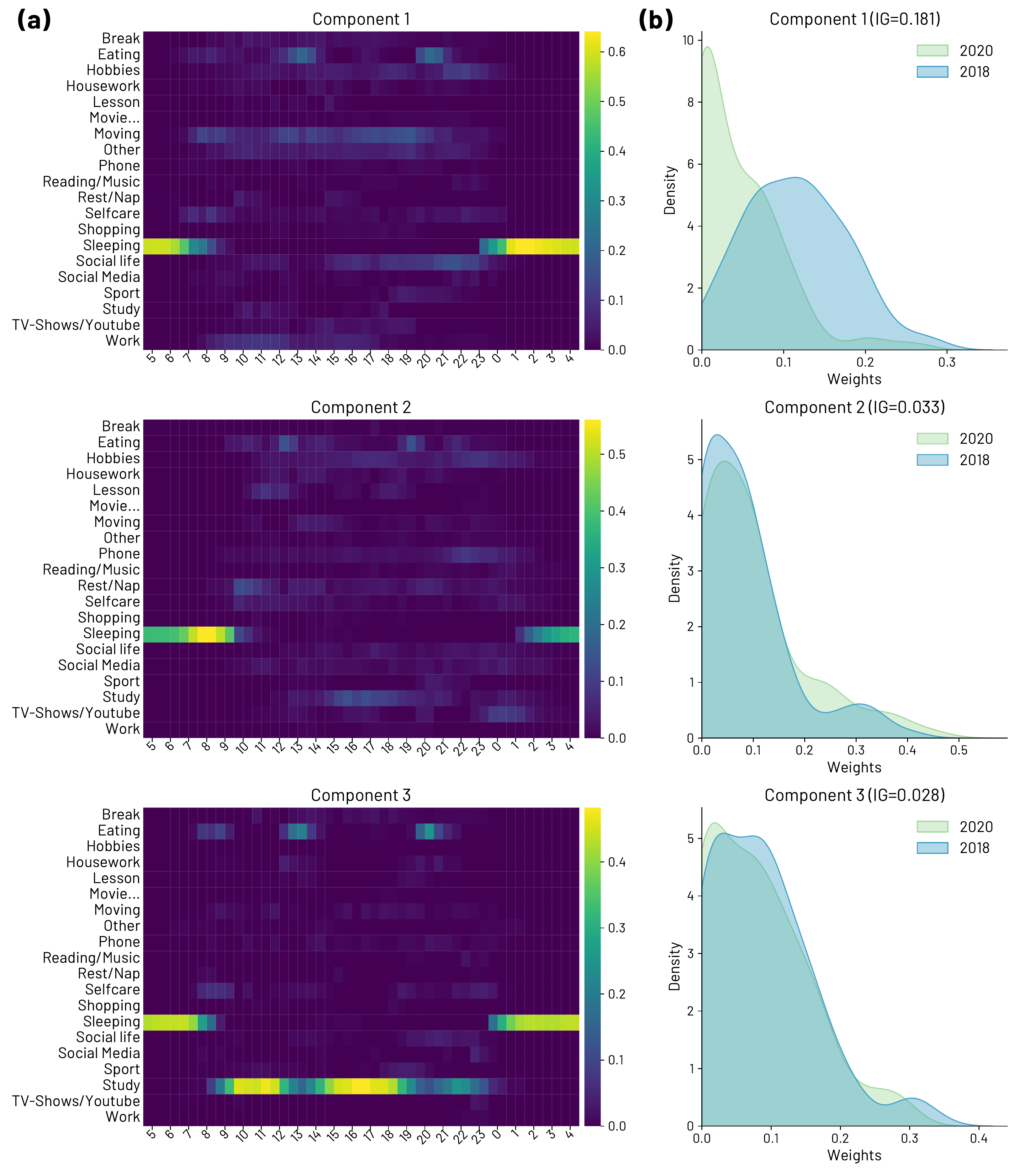}
    \caption{\csentence{Activity Routines 1-3.} In \textbf{(a)} we can see the components 1-3 extracted by the NMF algorithm for the \textit{activity} dimension. The vectors representing each component in the matrix $K$ have been visually rearranged in a matrix format, so that the rows represent the different possible activities, while columns are the 48 bins in which the day has been divided (i.e., every 30 minutes). Thus, the heatmaps shows how present an activity is in a specific time slot. In \textbf{(b)} we can see the distributions of the weights each student (divided by year) has for the components, showing how much that routine is important for the student's behaviour. We see that only the first component has different distributions for 2018 and 2020. The Information Gain (IG) shown quantifies how much knowing the component can tell apart 2018 behaviours from 2020 behaviours, and thus quantifies the weights distributions difference.}
    \label{fig:activity_components1}
\end{figure}

\begin{figure}[ht!]
    \includegraphics[width=.98\textwidth]{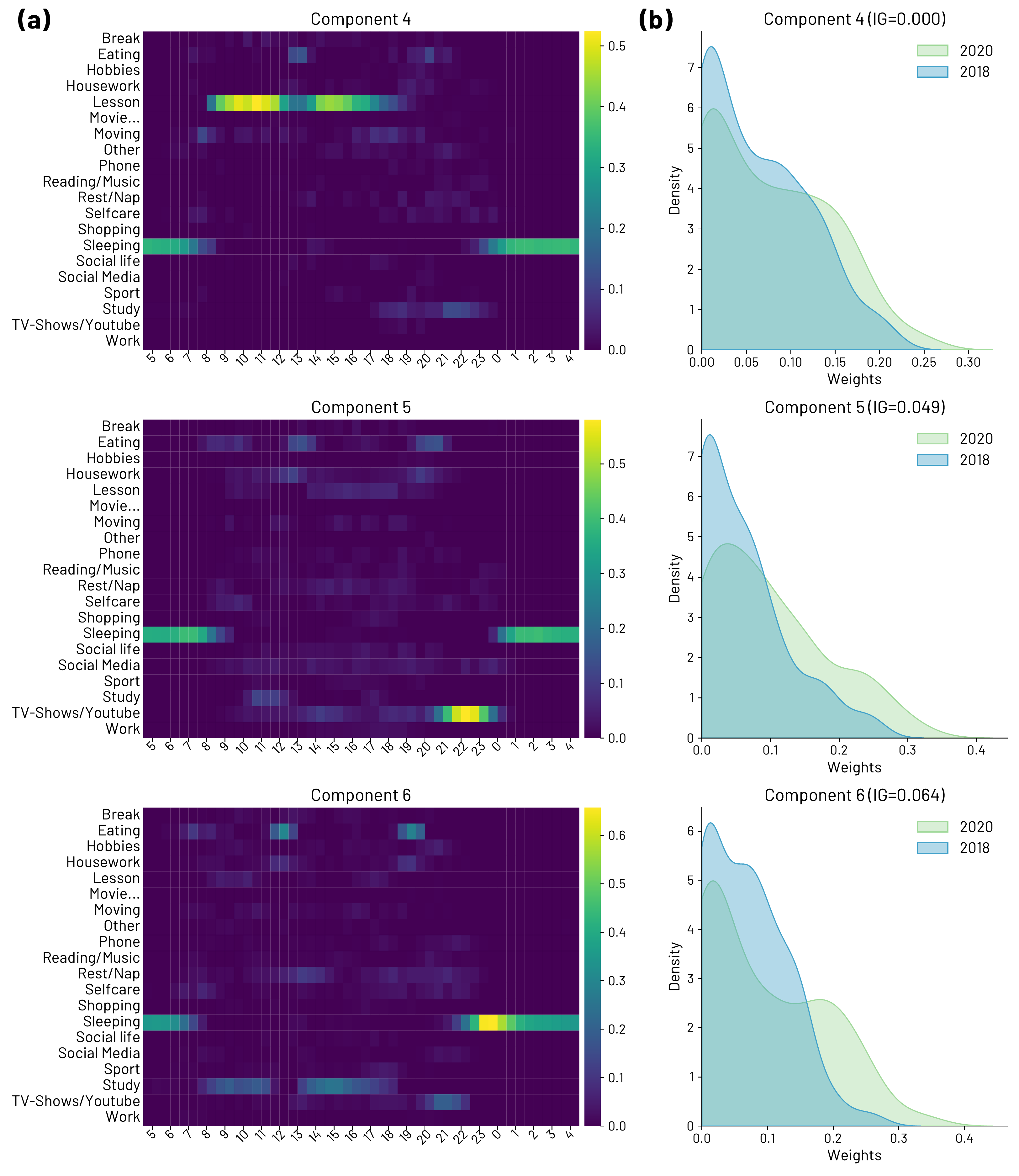} 
    \caption{\csentence{Activity Routines 4-6.} Similar to \Cref{fig:activity_components1}, \textbf{(a)} shows the NMF extracted components 4-6 for the \textit{activity} dimension, while \textbf{(b)} shows the students' weights distribution associated with them. We see that in this case the distributions do not differ much between 2018 and 2020. In fact, the IG is very low for each component.}
    \label{fig:activity_components2}
\end{figure}

In \Cref{fig:activity_components1} and \Cref{fig:activity_components2}, we observe the 6 \textit{activity} components (or routines) extracted by the algorithm. Note that the order of the components is not related to their significance, unlike other techniques such as Principal Component Analysis. The heatmaps (\Cref{fig:activity_components1}(a) and \Cref{fig:activity_components2}(a)) illustrate the weight of each activity at specific times of the day, with higher weights indicating a higher occurrence of the activity. As it is possible to observe, during the day, certain routine patterns are more prominent than others, with essential activities like eating and sleeping being influenced by other activities of the day. For example, this can be seen in Component 5, where students go to sleep later than other components (e.g., Component 6) because they watch YouTube and TV shows. Furthermore, NMF correctly extracts routines related to a student's academic life, and activities such as attending lessons and studying exhibit high weights in three components (3, 4, and 6). However, other behavioural patterns are less clear. Components 1 and 2 (in \Cref{fig:activity_components1}(a)) include several activities that share time allocations, indicating more complex daily routines. Component 1 is the routine that mostly captures social life, with higher weights for commuting (Moving) and social activities during the late afternoon and evening (Social Life).

A natural question that follows after looking at the most dominant \textit{activity} routines extracted by the NMF algorithm, is whether the behaviour described by a component is more associated with typical behaviours of students in 2018 or 2020. We can investigate such association by looking at the weights extracted by the NMF algorithm present in matrix $W$ (see \Cref{subsec:nmf}).

The weights assigned to students on the extracted components provide a measure of the presence of each component in their behaviour. Higher weights indicate a stronger association with the respective NMF component. \Cref{fig:activity_components1}(b) and \Cref{fig:activity_components2}(b) display the distributions of component weights for students in 2018 and 2020. A noticeable difference is observed in the first component, characterized by students moving around and engaging in evening social activities, which is more closely associated with a behaviour typical of 2018, as indicated by the higher values of the weight distribution. The distribution for students in 2020 is much more skewed towards 0, meaning that this kind of behaviour was not present, probably due to the COVID-19 restrictions imposed during 2020 (see \Cref{sec:data}). Also, Component 6 (\Cref{fig:activity_components2}) displays a (small) distinction between the 2018 and 2020 students groups, with typical behaviours in which students study during the day and watch YouTube, TV shows, or similar activities in the evening. Here, the weight distribution for students in 2020 exhibits a heavier tail and a second saddle point, with larger values compared to the weights of 2018 students, thus associating this more sedentary routine more with 2020 than 2018.

Examining the weight distributions of the other components, it becomes evident that there are just small differences between the two groups of students. This observation is quantitatively supported by the Information Gain (IG), computed for each of the components, which quantifies the amount of information that knowing the weight of a component gives in discriminating whether the described behaviour is typical of a student in 2018 or 2020 (higher IG values indicate a larger separation in the distributions). As reported in Figure \ref{fig:activity_components1}(b) and Figure \ref{fig:activity_components2}(b), we can see that the first component carries the most informative weight, while the remaining components do not exhibit notable differences between 2018 and 2020.

Interestingly, the extracted \textit{activity} routines displayed minimal changes between 2018 and 2020. Despite the pandemic restrictions, students in Trento maintained their activity behaviours focused on lessons and studying, while adapting their location and social interactions.

\begin{figure}[ht!]
    \includegraphics[width=.98\textwidth]{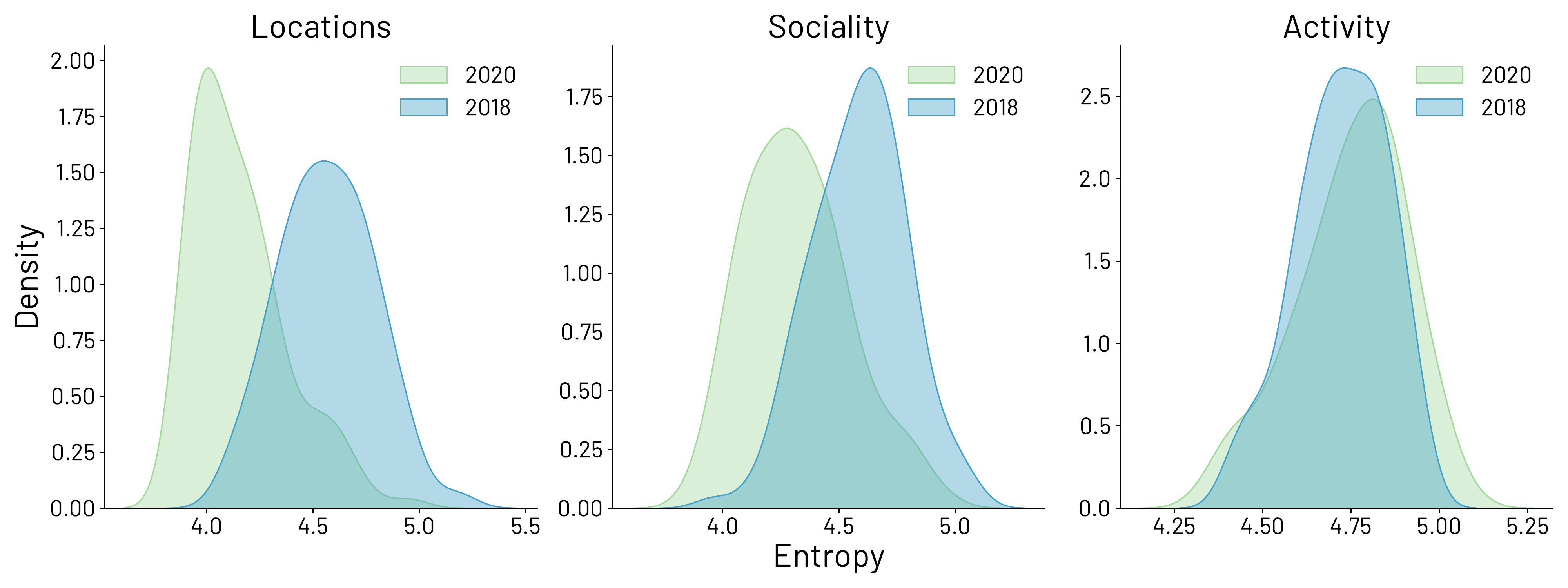}
    \caption{\csentence{Entropy for the Types of Information.} The three plots show the distribution of the entropy of 2018 and 2020 students' behaviour for \textit{location}, \textit{sociality} and \textit{activity} respectively. We see that the variety of activities remains the same, but students visited less diverse locations and stayed with less diverse companies during the COVID-19 pandemic.}
    \label{fig:entropy}
\end{figure}

As detailed in \Cref{app:locs_soc}, more pronounced differences were observed in the locations visited and in the social interactions between the two time periods. These disparities were expected due to the imposed restrictions, which limited mobility and reduced interpersonal interactions in 2020. Among the four \textit{location} routines identified through the NMF decomposition, staying at home throughout the day exhibited a significantly stronger association with students in 2020, while going to the university was less common in 2020 compared to 2018 (Components 1 and 4, see \Cref{appfig:locations}). These differences are supported by higher values of Information Gain.

Furthermore, the imposed restrictions also affected social gatherings, which, along with limitations on location visits, influenced students' social behaviour. Our analysis revealed differences in three out of seven extracted \textit{sociality} components. These three routines encompassed spending a day alone, spending a day with friends, and a working day with classmates (Components 5, 6, 7, see \Cref{appfig:sociality}). As expected, higher weights were observed for a day spent alone among 2020 students, while routines involving friends and classmates showed higher weights for 2018 students.

To support our findings, we conducted a quantitative evaluation of the diversity of behaviours between 2018 and 2020. To measure this diversity, we calculated the Shannon's entropy of students' behaviours using the \textit{activity}, \textit{location}, and \textit{sociality} matrices. The distributions depicted in \Cref{fig:entropy} represent the entropy computed for each student's normalized counts across the three types of information captured in the Time Diaries.

The results clearly indicate that in the \textit{activity} dimension, the diversity of activities performed by students in 2018 and 2020 is similar, suggesting that the variety of activities did not decrease significantly in 2020. However, in the \textit{location}, and \textit{sociality} dimensions, the entropy for 2020 is lower than the entropy of 2018. This indicates that the COVID-19 restrictions had a noticeable impact on the diversity of behaviours in these dimensions, with a reduction in the range of locations visited and the social interactions engaged in by students in 2020. The lower entropy values suggest a more limited range of options and routines in terms of locations and social behaviours due to the pandemic restrictions.

\subsection{Exploring Variations in Student Routines: Where and With Whom}
\label{subsec:change}

In the previous section, we observed that despite the COVID-19 pandemic's government restrictions, there were minimal changes in the \textit{activities} performed by students in 2020 compared to those in 2018. However, as expected, the \textit{location} and \textit{sociality} dimensions of their behaviour exhibited larger variations. This observation raised the question of how these two dimensions changed in conjunction with the activities, specifically where and with whom these activities were performed.

To gain insights into these variations, we applied the NMF decomposition once again, but this time we concatenated the behavioural matrices. Initially, to examine changes in locations for the activities, we concatenated the \textit{activity} behaviour matrix with the \textit{location} behaviour matrix. Subsequently, we explored the social dimension of activities by concatenating the \textit{activity} and \textit{sociality} behaviour matrices. Additionally, to enable a more focused analysis, we separated the groups of students in 2018 and 2020. By doing so, the NMF algorithm can extract the most dominant behaviours specific to each group. This approach allowed us to look into routines specific to each year and subsequently compare them, to potentially reveal shifts in \textit{location} and \textit{sociality} of the activities.

\begin{figure}[ht!]
    \includegraphics[width=.99\textwidth]{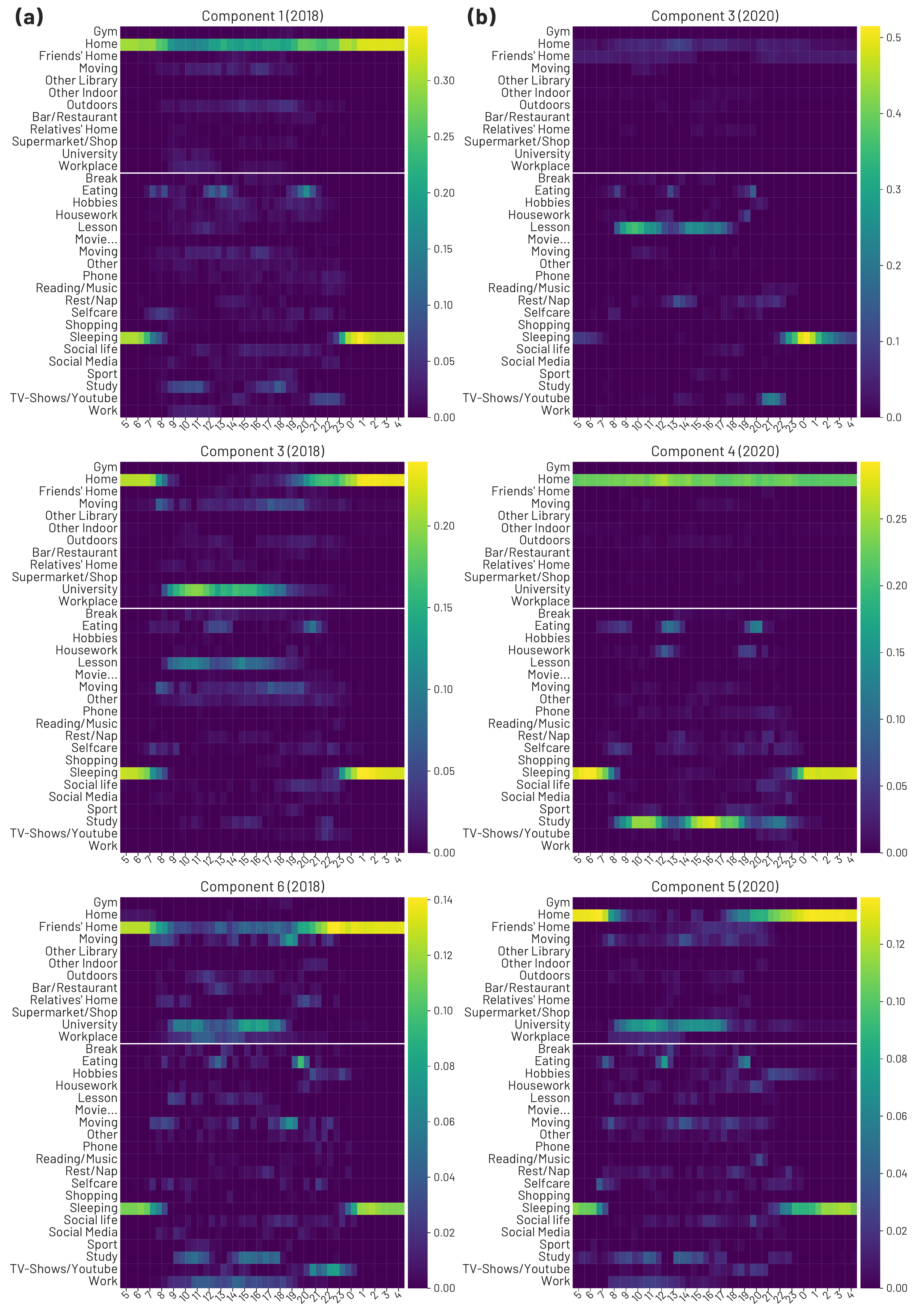}  
    \caption{\csentence{Location and Activity Joint Components.} This figure shows the joint \textit{location} and \textit{activity} (separated by the white line) components that most differentiate students behaviour before and during the pandemic. \textbf{(a)} includes routines from 2018, while \textbf{(b)} includes routines from 2020. We can see the shift from conducting activities such as studying and having lesson from the university to private homes}
    \label{fig:loc_act}
\end{figure}

\subsubsection{Where: Activities and Locations}
\label{subsubsec:act_loc}

We begin our exploration of the variation in students' routines by examining the relationship between activities and their respective locations. Firstly, we observe that the optimal number of components of the NMF algorithm, determined using the cophenetic coefficient, differs between students in 2018 and 2020, with values of 7 and 5, respectively. This difference suggests that the extracted behaviour in 2018 exhibits greater diversity compared to 2020, which aligns with our previous findings regarding the entropy of students in the \textit{location} behavioural matrix, where the diversity of locations is lower in 2020 (\Cref{subsec:stud_behav}). In 2020, all \textit{activity} routines predominantly occur either at home or at relatives' homes except Component 6, which includes days at the university and Component 3, which does not retain much information about the location. Conversely, in 2018, we observe that \textit{activity} routines, in addition to the home or at relatives' homes, took place at university, friends' homes, and even workplaces (Components 3 and 6).

The differentiation in students' behaviour can be observed in \Cref{fig:loc_act} (the complete components can be found in \Cref{subapp:act_loc}). When examining the activities, we find that the main routines extracted by NMF for both groups of students primarily revolve around activities such as lessons and studying. In 2018 (\Cref{fig:loc_act}(a)), these activities are closely associated with being at university (as seen in Components 3 and 6). When, instead, students are at home (either alone or with their relatives), studying and having lesson are not as present, and their routine is more varied. Conversely, in 2020 students carry out the majority of their activities, including lessons and studying, at home. It is interesting that in 2020, there is still a component associated with being at the university, namely Component 6. In fact, it was still possible for students to go to the library and follow lessons in-person (if they were of the first year). Moreover, we see that when students go to the university, their routine gets more complicated and they do more activities.

There were also other minor differences, such as having social life being a unique activity in 2018, while housework is more distinctive among students in 2020, further highlighting the impact of the pandemic (see \Cref{appfig:act_loc1}, \Cref{appfig:act_loc2}). These results indicate that while the main activities remained consistent between the two groups of students, the pandemic restriction forced these activities to be done in different locations.

\subsubsection{With Whom: Activities and Sociality}
\label{subsubsec:act_soc}

\begin{figure}[ht!]
    \includegraphics[width=.99\textwidth]{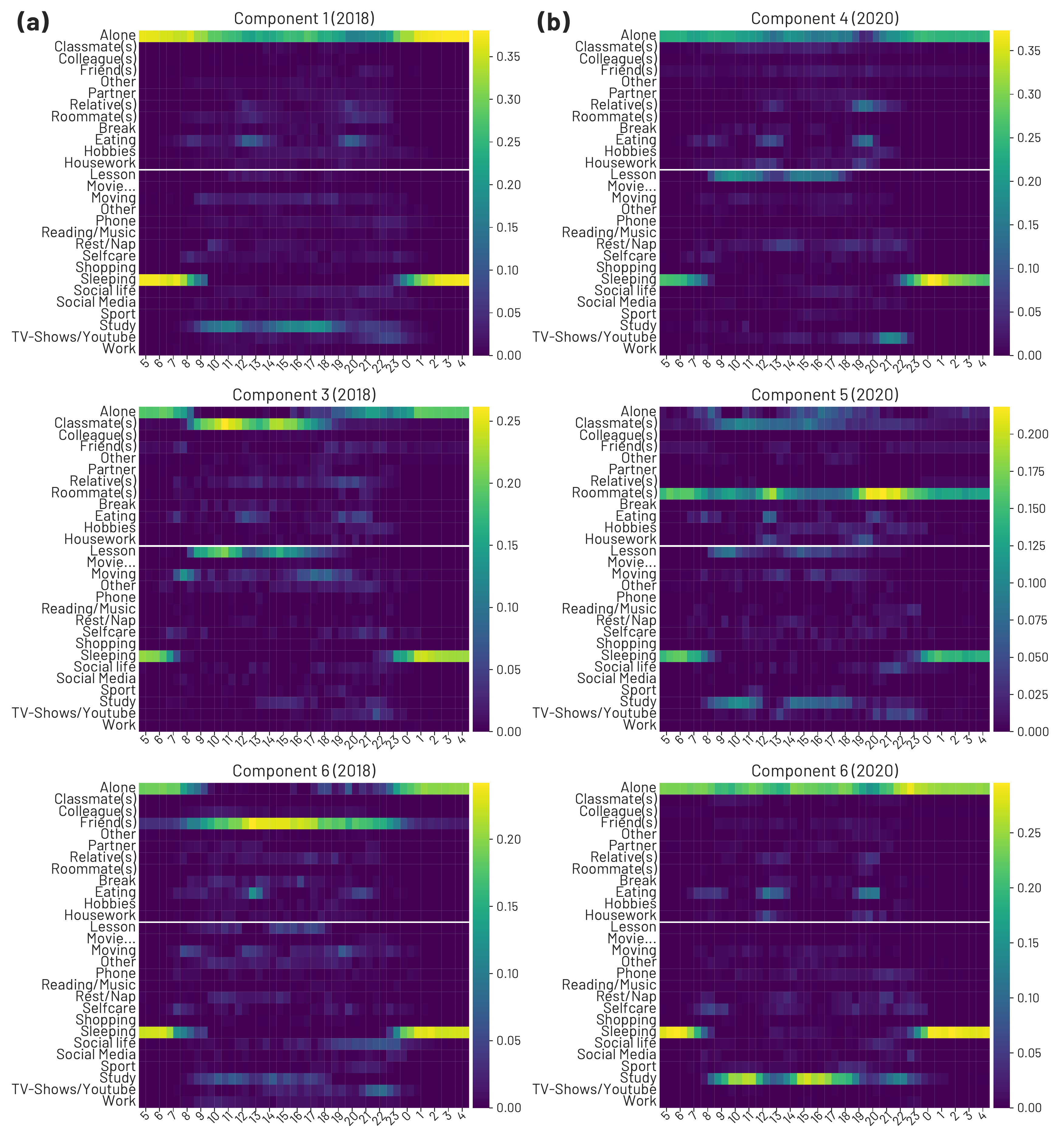}  
    \caption{\csentence{Sociality and Activity Joint Components.} This figure shows the joint \textit{sociality} and \textit{activity} components that most differentiate students behaviour before and during the pandemic. \textbf{(a)} includes routines from 2018, while \textbf{(b)} includes routines from 2020. The main activities are still \textit{Lesson} and \textit{Study}, which are social activities during 2018, while in 2020 students do them mostly alone (or with people in the same house)}
    \label{fig:soc_act}
\end{figure}

Turning our attention to the social component of activities, we observe that in 2020, the majority of the extracted components primarily involve activities performed alone, see \Cref{fig:soc_act}(b) (the complete components can be found in \Cref{subapp:act_soc}). Alternatively, these components involve a small number of individuals who likely reside in the same household, such as relatives, partners, or roommates. This finding is consistent with the COVID-19 restrictions that limited the variety of social interactions during that period.

In contrast, as shown in \Cref{fig:soc_act}(a), the social component of activities in 2018 exhibits a greater variation, which involved also interactions with friends and classmates (e.g., Components 3 and 6), while studying is the only activity strongly associated with being alone (Component 1).

Although the NMF algorithm extracted six dominant components for both years, in 2018, there is a greater diversity of activities compared to students in 2020, where the weights in the heatmaps are more concentrated on specific activities. The activity patterns observed in the resulting routines are similar to those detected when considering the combined \textit{locations} and \textit{activities} matrices, which again involve as main activities studying and having lessons. 

Notably, in 2018, there is a presence of mobility activities (e.g., \textit{Moving}), which are almost absent in 2020. Additionally, in Component 5 of 2020 (\Cref{fig:soc_act}(b)), we can observe the presence of social activities in the evening, primarily involving interactions with roommates. In contrast, in 2018, without government restrictions, social activities are more consistently carried out in the company of friends.

\section{Discussion and Conclusions}
\label{sec:disc}

In this work, we have analyzed the changes in the behavioural routines of students of the University of Trento due to the COVID-19 pandemic, focusing on three dimensions of behaviour: their \textit{activity}, their \textit{location} and their \textit{sociality}. Few studies have considered students' behaviour, and generally, they tend to examine just one facet of behaviour \cite{students_life}. Moreover, existing studies are often limited to the analysis of data that comes from either survey, where data is collected manually, or mobile phones, where data is generally passively collected when triggered by specific events (e.g., phone calls, app usage, etc.). By leveraging an Experience Sampling Method (ESM) approach on smartphones, we collected self-reported information on the daily activities of university students by means of Time Diaries, which allowed for high-resolution and multimodal experience sampling data collection in a real-world setting, with a high-level of control over participants' responses.

To investigate how students' routines and behaviours have changed during the COVID-19 pandemic, we applied Non-Negative Matrix Factorization (NMF). Thanks to its interpretability and clustering properties, this technique is particularly useful in identifying repeating patterns in the daily behaviours of people, both at an individual and a collective level. It has been successfully used to study behavioural routines in different contexts such as urban mobility \cite{nmf_latent}, people's chronotypes \cite{nmf_rythms} and interactions \cite{gauvin2014detecting}. Our findings suggest that despite the COVID-19 restrictions, which have significantly altered various aspects of our lives \cite{living_pandemic, urban_behav, adams2020inequality, van2020covid, wellbeing}, there is surprisingly little change in the activities performed by students between 2018 (pre-pandemic) and 2020 (during the pandemic). Additionally, as shown in our analysis, the diversity of activities did not change significantly.

However, leveraging the richness of the data at our disposal, which enables a comprehensive multimodal analysis of students' behaviour, we discovered that their adaptation to the pandemic circumstances primarily occurred in the \textit{location} and \textit{sociality} dimensions. We observed differences in these dimensions for dominant activities that constitute typical student routines, such as studying and attending classes. For these routines, our results indicate that, while the main activities remained consistent between the two groups of students, the pandemic restriction forced these activities to be done in different locations, and, thus, with different people. 

More in detail, in 2018, students' activities such as studying and having lessons were often located at the university. In contrast, in 2020, due to the COVID-19 pandemic, students predominantly engaged in activities, including studying and lessons, from home, although there was the possibility for them to follow lessons at the university. Notably, when students went to university, their routines became more complex and involved a broader range of activities. From a sociality point of view, in 2020, the majority of social activities were performed alone or with a small number of individuals from the same household, reflecting the impact of COVID-19 restrictions on limiting social interactions. In contrast, the \textit{sociality} components of activities in 2018 exhibited greater variation, including interactions with friends and classmates, indicating a broader range of social engagements before the pandemic.

While our results stem from data collected from two living lab initiatives, in which the sample of students is different for 2018 and 2020, we believe that the outcomes are still reliable. In fact, this limitation is mitigated by the fact that the samples are from similar populations (students), and the data is collected in similar settings: they attend the same university and faculty in the same city.

Concluding, in applying Non-Negative Matrix Factorization, we have shown its usefulness in modelling students' behaviour and in extracting meaningful patterns, i.e., routines. We believe that our contribution can be valuable for further understanding the impact of the COVID-19 pandemic on students. Moreover, we show that using a multimodal approach can be useful in identifying patterns of joint behavioural dimensions, and, thus, be used to understand the correlations among these dimensions. Finally, this methodology can also help correlate the behavioural dimensions with other facets of human activity, enriching the literature on well-being \cite{students_life}, stress levels \cite{students_stress}, and mood \cite{generalization} of students who were significantly affected by the pandemic.


\begin{backmatter}

\section*{List of Abbreviations}
    \begin{itemize}
        \item \textbf{ESM:} Experience Sampling Method
        \item \textbf{NMF:} Non-Negative Matrix Factorization
        \item \textbf{PCA:} Principal Component Analysis
        \item \textbf{IG:} Information Gain
    \end{itemize}

\section*{Availability of data and materials}
    The datasets generated and analysed during the current study are not publicly available due to GDPR regulation and pending full approval from UniTN DPO but may be available from Fausto Giunchiglia (\href{mailto:fausto.giunchiglia@unitn.it}{fausto.giunchiglia@unitn.it}).

\section*{Competing interests}
  The authors declare that they have no competing interests.

\section*{Funding}
    BL and SC are partly supported by the PNRR ICSC National Research Centre for High Performance Computing, Big Data and Quantum Computing (CN00000013), under the NRRP MUR program funded by the NextGenerationEU. BL and SC received also support from the project “AIAAA - Applicazioni di Intelligenza Artificiale e Analisi Avanzata", funded by the Autonomous Province of Trento.
    
    The research by IB and FG has received funding from the European Union's Horizon 2020 FET Proactive project “WeNet – The Internet of us”, grant agreement No 823783.

\section*{Author's contributions}
    NAG, SC and BL designed the research. IB and FG designed the data collection experiments and provided the data after cleaning and preparation. NAG processed the data and performed the analysis. All the authors contributed to the writing of the article. All authors read and approved the final manuscript.

\section*{Acknowledgements}
  Not applicable
  

\bibliographystyle{bmc-mathphys} 
\bibliography{ref}
\nocite{label}

\end{backmatter}


\clearpage
\appendix

\renewcommand{\thefigure}{A\arabic{figure}}
\renewcommand{\thetable}{A\arabic{table}}
\setcounter{figure}{0}

\section{Data Mapping}
\label{app:mapping}

Because of the refinement in the WeNet study, a mapping is needed to make the two datasets comparable. \Cref{tab:activity_map} shows the mapping for the \textit{activities}, while \Cref{tab:locations_map} contains the \textit{locations} mapping. For \textit{activities}, the SmartUniTn (a bit different from the cited paper, as there were refinements from 2017 to 2018) is on the left and it is the one used for the analysis. Note that for the mobility part, all the possibilities have been further mapped to \textit{Moving}. For the \textit{location} data, we also aggregated some very specific locations to make spotting patterns an easier task.

\begin{table}[h!]
    \centering
    \caption{Mapping of Activity}
    \label{tab:activity_map}
    \begin{tabular}[width=\textwidth]{l|l}
        \hline
        \textbf{SmartUniTn} & \textbf{WeNet} \\ \hline
        Sleeping & (1) Sleeping; (41) I will go to sleep \\ \hline
        Selfcare & (2) Personal care \\ \hline
        Eating & (3) Eating \\ \hline
        Study & (5) Study/work group; (38) I have a work/study meeting \\ \hline
        Lesson & \makecell[l]{(6) Lecture/seminar/conference/university meeting; \\ (40) I am starting classes/lessons/lab} \\ \hline
        Social life & (12) Social life; (13) Happy Hour/Drinking/Party \\ \hline
        Watching Youtube Tv-shows etc. & (18) Watching TV, video, YouTube, etc. \\ \hline
        Social media (Facebook Instagram etc.) & \makecell[l]{(16) Surfed or seeking via Internet; \\ (17) Social media (Facebook Instagram etc.)} \\ \hline
        Coffee break cigarette beer etc. & (9) Break (coffee, cigarette, drink, etc.) \\ \hline
        All the phone; in chat WhatsApp & \makecell[l]{(14) Phone/Video calling; \\ (15) In chat on Internet or reading, sending e-mail;} \\ \hline
        Reading a book; listening to music & \makecell[l]{(19) Listening to music; (20) Reading a book, periodicals, news, etc.} \\ \hline
        \makecell[l]{Movie Theater Theater Concert Exhibit \\ Entertainment and Culture} & \makecell[l]{(21) Movie Theatre Concert ... ; \\ (22) Entertainment Exhibit, and Culture; \\ (23) Others Entertainment and Culture; \\ (39) I am at the cinema/theater/hospital/church} \\ \hline
        Housework & \makecell[l]{(4) Cooking, Food preparation \& management; \\ (30) Household and family care} \\ \hline
        Shopping & (31) Grocery Shopping; (32) Other Shopping \\ \hline
        Sport & (28) Sport; (37) I will participate in sports activities \\ \hline
        Rest/nap & (7) Did not do anything special; (8) Rest/nap \\ \hline
        Hobbies & (27) Free Time Study; (24) Arts; (25) Hobbies; (26) Games \\ \hline
        Work & (33) Work \\ \hline
        By car & (48) by car; (49) by car as passenger \\ \hline
        By foot & (10) Walking; (42) on foot \\ \hline
        By bike & (43) by bike \\ \hline
        By bus & \makecell[l]{(11)Travelling (44)by bus/tram; (45)by metro/subway/underground} \\ \hline
        By train & (46) by train \\ \hline
        By motorbike & (47) by motorbike \\ \hline
        Other & \makecell[l]{(34)Other (36)Others; (29)Voluntary work, and participatory activities}
    \end{tabular}
\end{table}

\begin{table}[h!]
    \centering
    \caption{Mapping of Locations}
    \label{tab:locations_map}
    \begin{tabular}[width=\textwidth]{l|l|l}
        \hline
        \textbf{Final Mapping} & \textbf{SmartUniTn} & \textbf{WeNet} \\ \hline
        Gym & \makecell[l]{Gym, swimming pool, Sports centre...} & \makecell[l]{Gym, swimming pool, Sports centre...} \\ \hline
        Home & \makecell[l]{Home apartment/room; \\ Home garden/patio/courtyard} & \makecell[l]{Home apartment/room; \\ Home garden/patio/courtyard} \\ \hline
        Friends' Home & \makecell[l]{House (friends' others')} & \makecell[l]{House (friends' others')} \\ \hline
        Moving & \makecell[l]{Moving... From.. To} & \makecell[l]{Moving... From.. To} \\ \hline
        Other Library & \makecell[l]{Other Library} & \makecell[l]{Other Library} \\ \hline
        Other Indoor & \makecell[l]{Movie Theatre Museum...} & \makecell[l]{Movie Theatre Museum...; Another indoor place} \\ \hline
        Outdoors & \makecell[l]{Outdoors} & \makecell[l]{In the street; Public Park/Garden; \\ Countryside/mountain/hill/beach; \\ Another outdoor place; \\ Hotel, guesthouse, camping site} \\ \hline
        Bar/Restaurant & \makecell[l]{Café, pub, bar; \\ Restaurant, pizzeria, Street food vendor} & \makecell[l]{Café, pub, bar; \\ Restaurant, pizzeria, Street food vendor} \\ \hline
        Relatives' Home & \makecell[l]{Relatives' Home} & \makecell[l]{Relatives' Home; \\ Weekend home or holiday apartment} \\ \hline
        Supermarket/Shop & \makecell[l]{Grocery Shop; Shops, shopping centres; \\ Supermarket; Street markets} & \makecell[l]{Grocery Shop; Shops, shopping centres, indoor \\ markets, other shops; Supermarket; Street markets} \\ \hline
        University & \makecell[l]{Canteen; Classroom/Study hall; \\ Classroom/Laboratory; University Library; \\ Other university place} & \makecell[l]{Canteen; Classroom/Study hall; \\ Classroom/Laboratory; University Library; \\ Other university place} \\ \hline
        Workplace & \makecell[l]{Workplace} & \makecell[l]{Workplace/office} \\
    \end{tabular}
\end{table}

\clearpage
\section{Location and Sociality Routines}
\label{app:locs_soc}

In this section we show the NMF extracted components for the \textit{location} and the \textit{sociality} behaviour dimensions.

\subsection{Locations}
\label{subapp:locs}

\begin{figure}[ht!]
    \includegraphics[width=.99\textwidth]{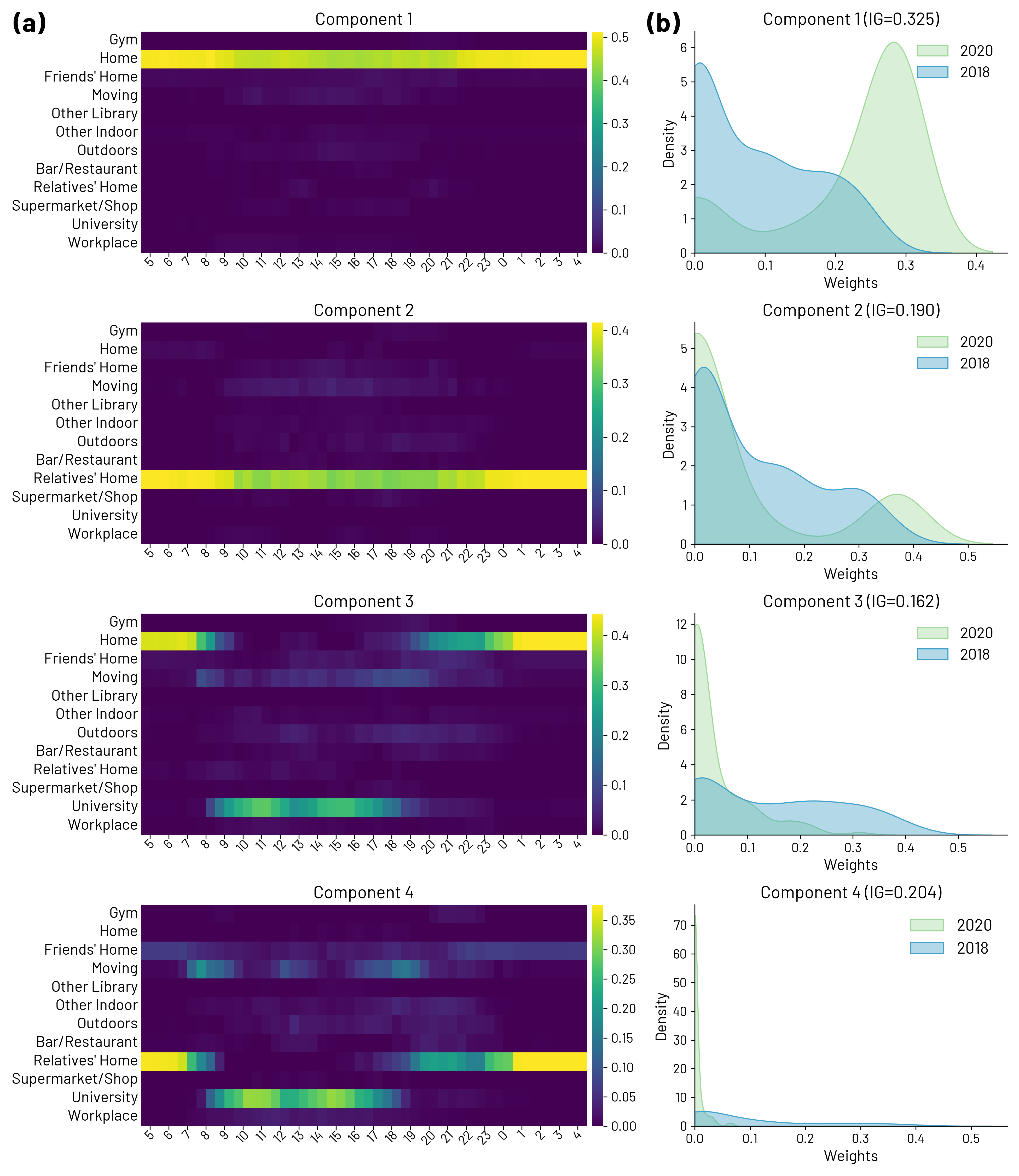}
    \caption{\csentence{Location Routines.} Like in the main manuscript, \textbf{(a)} shows the extracted components and \textbf{(b)} the associated weights for the students, distinguishing between the 2018 and 2020 populations. Here we see students of 2020 staying more at home and not going to the university}
    \label{appfig:locations}
\end{figure}

\Cref{appfig:locations} shows the 4 \textit{location} routines and their weights. We see how staying at home has higher weights for 2020, while going to the university is way more common in 2018. These results show the effect of the restrictions.

\subsection{Sociality}
\label{subapp:soc}

\begin{figure}[ht!]
    \includegraphics[width=.99\textwidth]{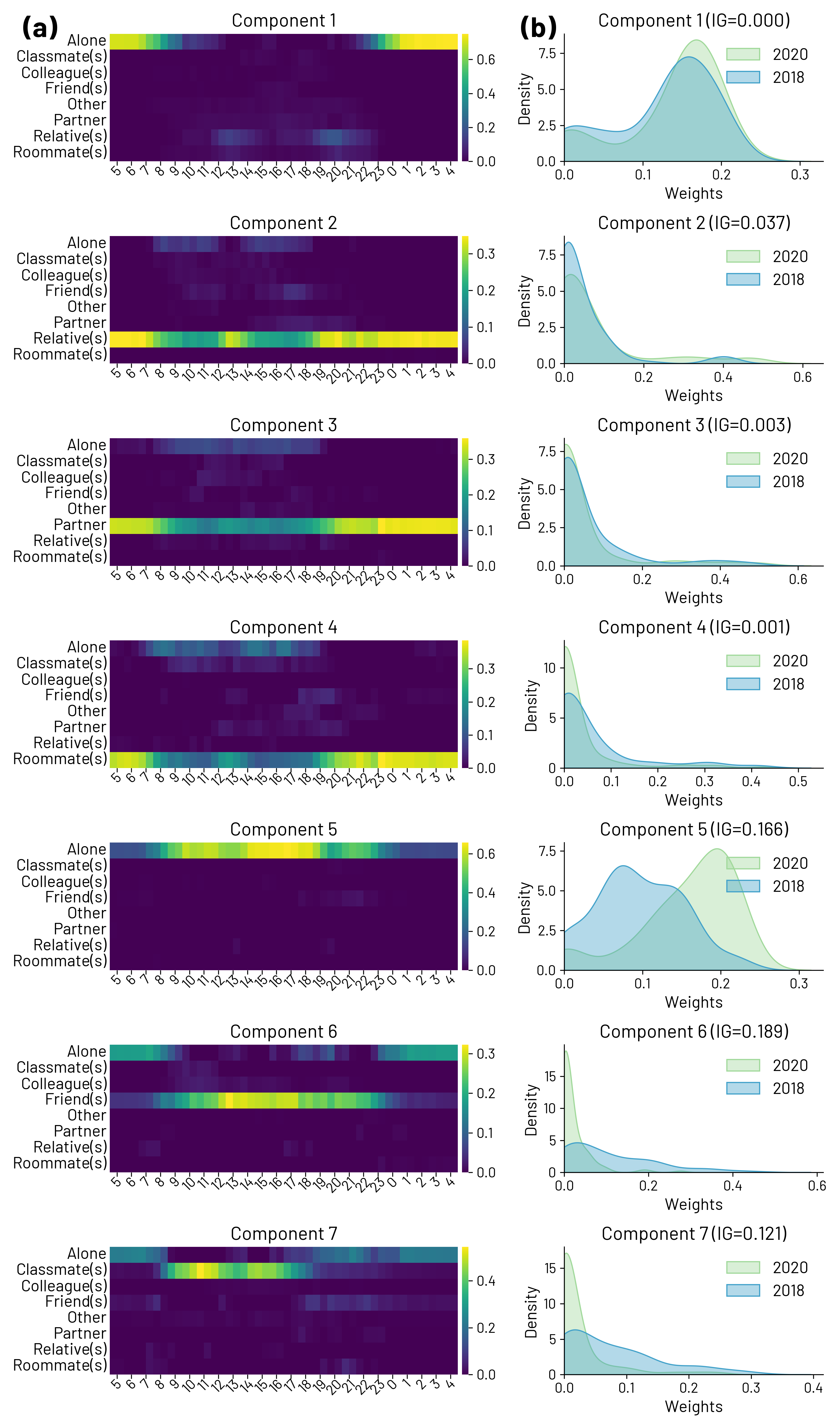}
    \caption{\csentence{Sociality Routines.} Like in the main manuscript, \textbf{(a)} shows the extracted components and \textbf{(b)} the associated weights. We see the shift towards more time spent alone and less with friends and classmates}
    \label{appfig:sociality}
\end{figure}

In \Cref{appfig:sociality} we see that some \textit{sociality} routines remain the same. Still, there are changes, in the fact that being alone is much more common in 2020 and, consequently, social gatherings with friends and classmates are very rare during.

\clearpage
\section{Joint Activities, Locations and Sociality}
\label{app:act_loc_soc}

In this section, we show the complete outcome of the NMF algorithm applied to the joint information of \textit{activities}, \textit{location} and \textit{sociality}.

\subsection{Activities with Locations}
\label{subapp:act_loc}

\begin{figure}[ht]
    \includegraphics[width=.99\textwidth]{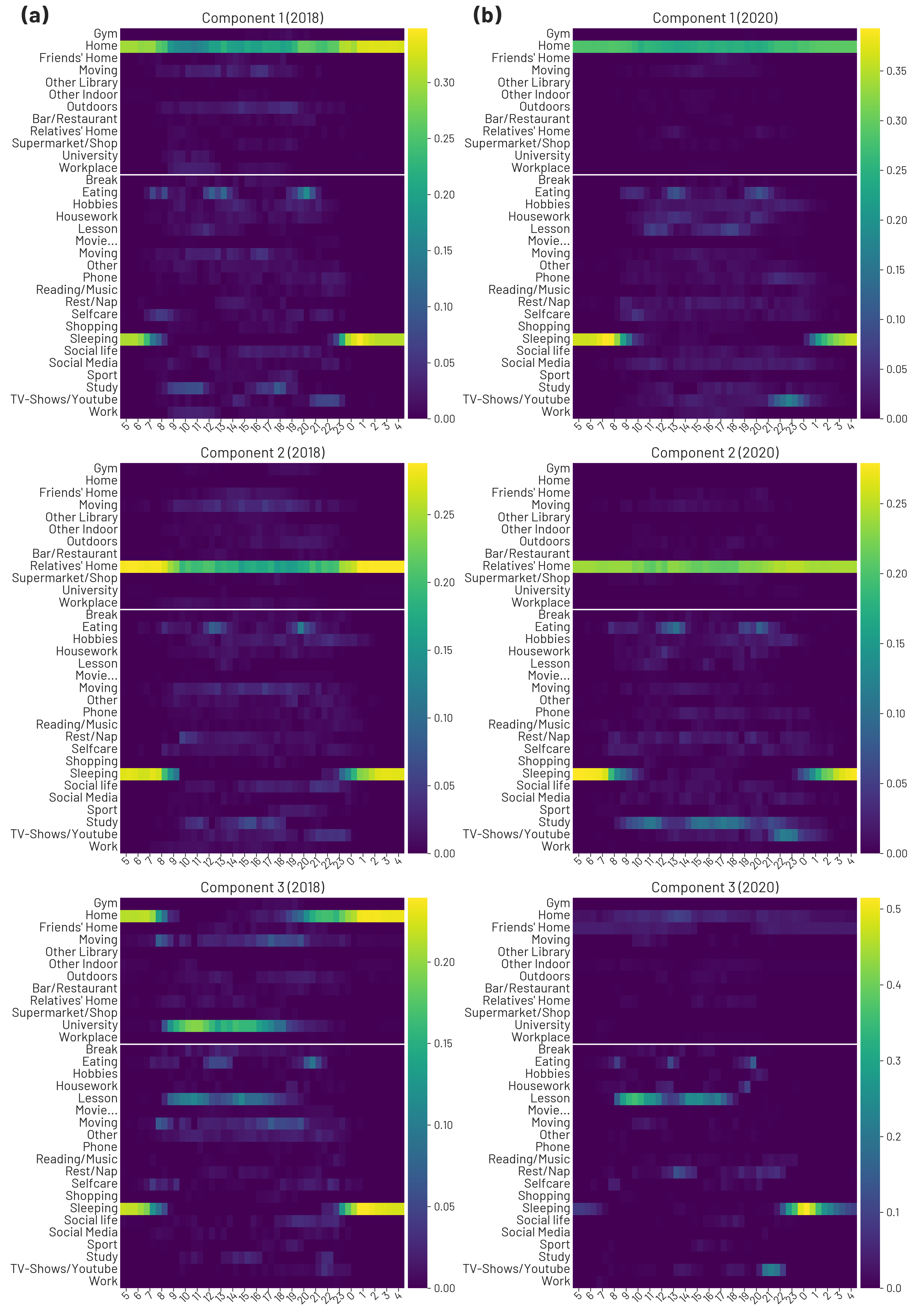}
    \caption{\csentence{Activities and Locations 1-3.} The components in \textbf{(a)} are relative to 2018, while the ones in \textbf{(b)} are relative to 2020. The main effect is a shift of study and lesson from the university to private homes}
    \label{appfig:act_loc1}
\end{figure}

\begin{figure}[ht]
    \includegraphics[width=.99\textwidth]{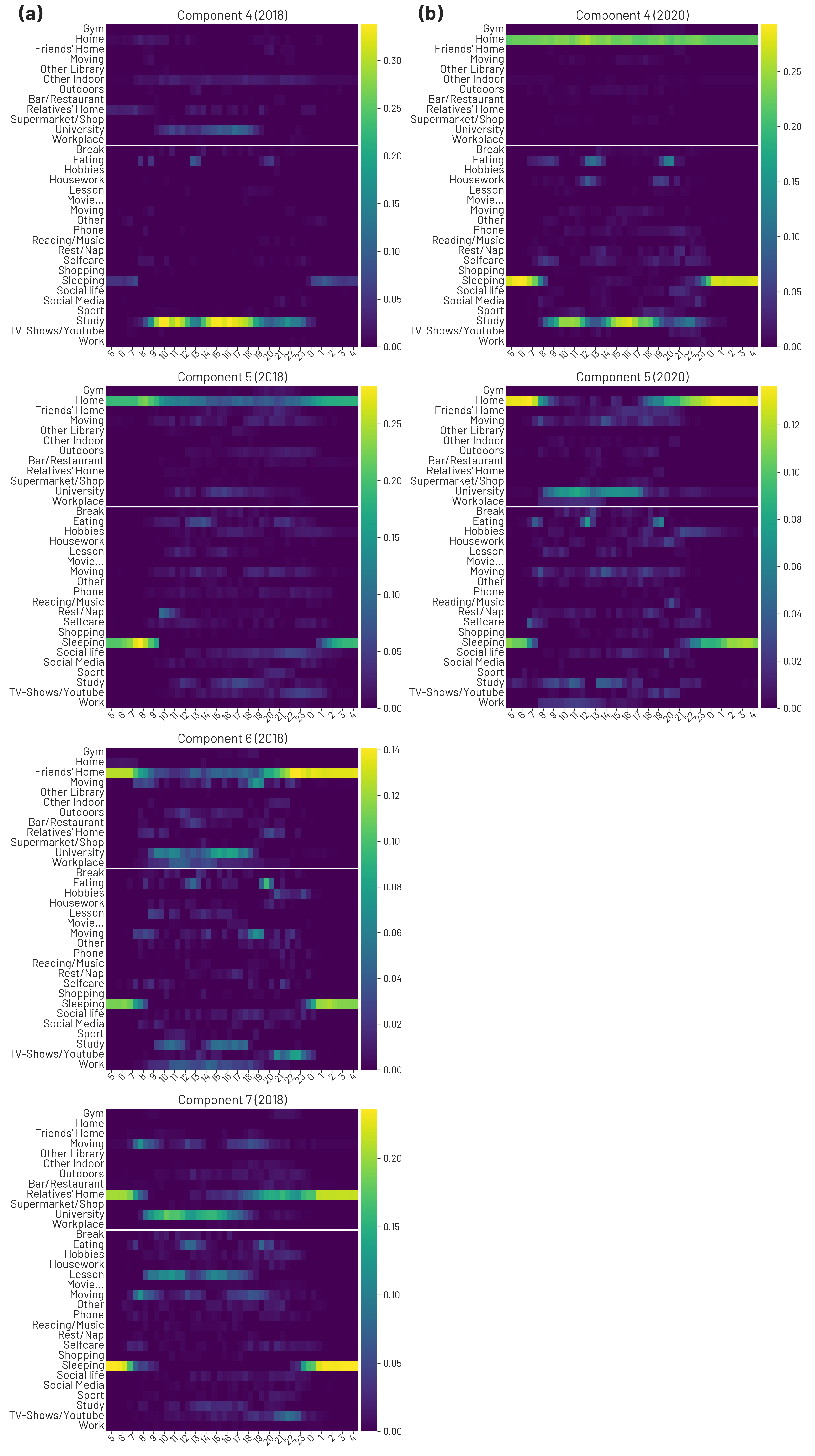}
    \caption{\csentence{Activities and Locations 4-7.} \textbf{(a)} shows 2018 components, while \textbf{(b)} shows 2020 components. Specific activities are done in universities in 2018, while in 2020 going there is an occasion for more diverse activities}
    \label{appfig:act_loc2}
\end{figure}

\Cref{appfig:act_loc1} and \Cref{appfig:act_loc2} show the components for the joint information of \textit{locations} and \textit{activities}. Like in the main text we see Component 4 of 2018 and Component 3 of 2020 without high weights for \textit{locations}: this because they are already encoded by the other components. The other components show how the study and lesson activity moved from the university to private homes. \textit{Activity} routines are very clear for 2020 when students are at home, while they vary more if they go to the university.

\subsection{Activities with Sociality}
\label{subapp:act_soc}

\begin{figure}[ht]
    \includegraphics[width=.99\textwidth]{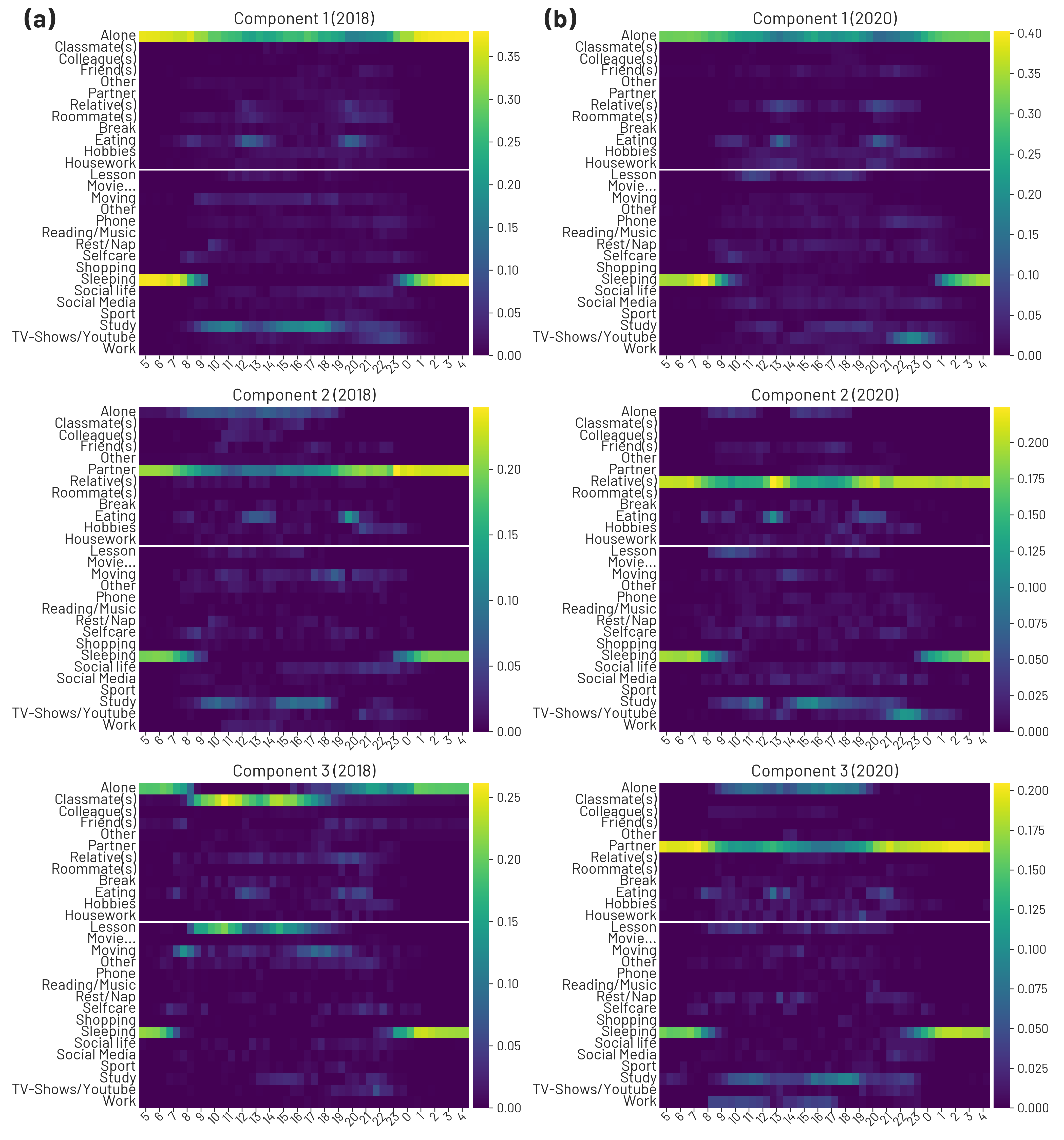}
    \caption{\csentence{Activities and Sociality 1-3.} The components in \textbf{(a)} are relative to 2018, while the ones in \textbf{(b)} are relative to 2020. We see that in 2020 social life is mostly alone or with smaller company}
    \label{appfig:act_soc1}
\end{figure}

\begin{figure}[ht]
    \includegraphics[width=.99\textwidth]{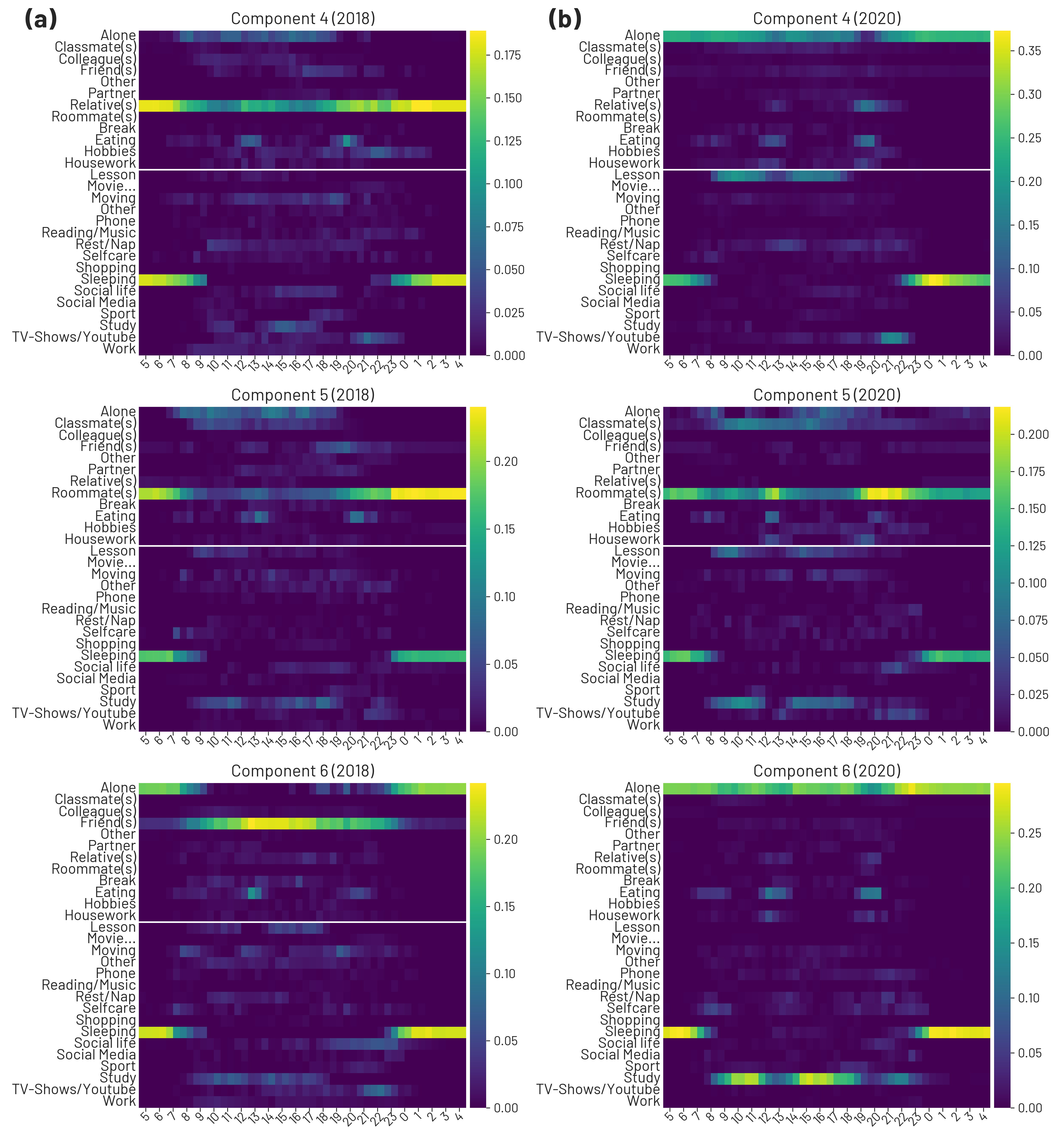}
    \caption{\csentence{Activities and Sociality 3-6.} The components in \textbf{(a)} are relative to 2018, while the ones in \textbf{(b)} are relative to 2020. In these components we a bigger focus on study for 2020 even with different companies}
    \label{appfig:act_soc2}
\end{figure}

The joint components of \textit{activity} and \textit{sociality} are shown in \Cref{appfig:act_soc1} and \Cref{appfig:act_soc2}. We see how there is almost no social activity large groups (i.e., classmates and friends) in 2020, and activities that were done with a bigger company in 2018 (i.e., study and having lesson) are done mostly alone. We also see the absence of social life done with friends in 2020, which is instead present in 2018, and social life is just with roommates.

\end{document}